\documentclass[12pt]{iopart}
\usepackage{iopams}
\usepackage{braket}
\usepackage{slashed} 
\usepackage{cite}

\begin{document}

\title{A manifestly Lorentz covariant, interacting and non-commutative Dirac equation}

\author{P~H~Williams$^1$ and F~G~Scholtz$^{1,2}$}
\address{$^1$Institute of Theoretical Physics, University of Stellenbosch, Stellenbosch 7600, South Africa}
\address{$^2$National Institute for Theoretical Physics (NITheP), Stellenbosch 7600, South Africa}

\begin{abstract}
We propose a manifestly Lorentz covariant, non-commutative Dirac equation for charged particles interacting with an electromagnetic field.  The equation is formulated on the operator level, but operators are not composed through the normal operator product, but a modified product that restores the Lorentz covariance.  This equation is solved for the free particle and a particle moving in a constant magnetic field.  An abstract action, constructed on the operator level, that yields this equation as equation of motion is also derived.  To relate this formalism to current formulations of non-commutative quantum field theories, this action is written in a coherent state basis, leading to an action in 4-dimensional Minkowski space-time.   The resulting action differs from existing non-commutative actions, but still exhibits non-commutative effects through non-locality.
\end{abstract}

%
\vspace{2pc}
\maketitle
\section{Introduction}

Non-commutative space-time has firmly established itself as a potential scenario for the quantum structure of space-time at short length scales \cite{Doplicher}.  Once this scenario has been adopted, it becomes necessary to revisit the formulation of quantum mechanics \cite{scholtz1} and quantum field theories \cite{doug} on these spaces.  Remarkably, very little modification of the standard formulation of quantum mechanics \cite{scholtz1} or quantum field theories \cite{doug} is required.  The main generalisation is that in the former wave-functions and the latter fields are replaced by elements of an associative algebra generated by the non-commutative coordinates.  Correspondingly the notion of a derivative and integral must be generalised to this setting, the former becoming a commutator and the latter a trace \cite{doug}.

However, for the simplest type of non-commutativity
\begin{equation}
\label{ncom}
\left[\hat{x}_\mu , \hat{x}_\nu \right] = i \theta_{\mu \nu},
\end{equation}
with $\theta_{\mu \nu}$ constants, rotational invariance is broken in 3-dimensional quantum theories and Lorentz invariance in 3+1-dimensional quantum field theories.  In addition these field theories have the undesirable property of UV/IR mixing, which may jeopardise their perturbative renormalisability \cite{doug}.  

In the case of 3-dimensional quantum mechanics the breaking of rotational symmetry can be overcome by adopting fuzzy sphere commutation relations \cite{scholtz2}.  In the case of 3+1-dimensional quantum field theories, it was more recently shown that the Lorentz symmetry is restored upon twisting \cite{chaichian}.  Despite this insight several outstanding and controversial issues still plague the twisted implementation of the Lorentz symmetry \cite{pinzul}.  The first difficulty one encounters is to carry out the standard Noether analysis and to identify conserved charges for the twisted Lorentz symmetry \cite{pinzul}.   Another obstacle is the quantisation of these theories, i.e., one can adopt the point of view that the standard quantisation procedure applies or one can also deform the canonical commutation relations (see \cite{pinzul} and references therein).  On the level of the functional integral this amounts to altering the measure \cite{pinzul}.  This has rather drastic consequences such as the absence of UV/IR mixing.  Indeed, in  \cite{pinzul} it is argued that UV/IR mixing may be related to a quantum anomaly of the Lorentz group, which is very closely related to the choice of functional integral measure.  At this point there seems to be no consensus between these different points of view.

Given this situation, a formulation of non-commutative quantum field theories, based on the commutation relations (\ref{ncom}), for which the Lorentz symmetry is manifest seems highly desirable.  This paper is a first attempt in this direction.  Here we focus on the Dirac equation for a charged particle interacting with an electromagnetic field and attempt to write this equation in a manifestly covariant way on the operator level.  We then derive an action, also on the operator level, which is manifestly Lorentz invariant and that yields this as equation of motion.  Once this had been achieved, one can apply a symbol map \cite{doug} under which the operators become fields on some emerging space-time.  Since this map is well under control, the hope is that one can in this way learn how to quantise these theories and implement the Lorentz symmetry.  Here we do not carry out this second step explicitly, but rather focus on the covariant Dirac equation.  

The paper is organised as follows: Section \ref{Formulation} briefly reviews the operator formulation of non-commutative quantum mechanics and field theories.  Section \ref{Lorentz Symmetry} reviews the derivation of the Lorentz generators in this formulation.  Section \ref{Dirac} introduces and solves the free Dirac equation.  It then proceeds to analyse the action of the Lorentz generators on the products of operators (fields) and establish a covariant form of the Dirac equation.  In addition the gauge and discrete symmetries of this covariant Dirac equation is also verified. Section \ref{Landau} solves the non-commutative Dirac equation in the presence of a constant background magnetic field, while Section \ref{action} introduces an action, on the operator level, that yields the covariant, interacting Dirac equation as equation of motion. This action is also rewritten as a non-local field theory using coherent states. Finally Section \ref{concl} discusses the results and draws conclusions.

\section{Formulation}
\label{Formulation}

We consider four dimensional non-commutative space-time with coordinate operators $\hat{x}^\mu$ that satisfy (\ref{ncom}).  The standard notion of fields on commutative space is now replaced with the notion of an associative algebra generated by the space-time operators  \cite{doug} and the action of the non-commutative quantum field theory is a real valued functional on this algebra.  As it is rather difficult and inconvenient to carry this construction out completely on the abstract level, we prefer to introduce a concrete unitary realisation of this abstract algebraic construct.  The way we do this is to first introduce a concrete realisation of the coordinate algebra as hermitian operators acting on a Hilbert space ${\cal H}_c$, which we refer to as configuration space.  The associative algebra generated by the coordinates can then be thought of as an algebra of operators acting on ${\cal H}_c$.  This has the added advantage that this algebra has a natural topology and we can also define notions such as boundedness, involution and trace.  The latter is particularly  important as it plays the same role as integration in a commutative theory.   

The precise class of operators that one should consider in constructing this associative algebra is somewhat of an open question \cite{doug}. A natural choice is the $C^*$ algebra of bounded operators.  However, boundedness is not sufficient to ensure the existence of traces and, subsequently, the trace of commutators also do not need to vanish, which is equivalent to the presence of boundary terms in a commutative theory \cite{doug}.  One would therefore like to impose a stronger condition then boundedness on the fields.  In ordinary commutative quantum mechanics one requires square integrability of the wave-function.  The non-commutative analogue of this is the class of Hilbert-Schmidt operators and it is therefore quite natural to identify the Hilbert space of a non-commutative quantum system with the Hilbert space of Hilbert-Schmidt operators  \cite{scholtz1}.  In a quantum field theory, especially one in which time is also non-commutative,  this may seem too strict a condition to place on the fields as the trace effectively includes, in the commutative analogue, also an integral over time.  However, it turns out that one never really explicitly needs this condition, except when writing down an action where it is necessary to ensure the existence of the latter and the absence of boundary terms when deriving equations of motion.  In what follows we therefore require the fields to be Hilbert-Schmidt, which can be seen as a technical condition that ensures that all the mathematical computations we carry out are well defined.  In the treatment below, we therefore identify the associative algebra of fields with the Hilbert space of Hilbert-Schmidt operators.

The first step in the construction above is to find an explicit unitary realisation of the coordinate algebra (\ref{ncom}).  To do this we first perform a simple coordinate transformation (also found in \cite{sinha} for three dimensions). Consider a transformation $\hat{x}^\mu\rightarrow \hat{\tilde{x}}^\mu$ such that:
\begin{equation}
[\hat{\tilde{x}}^\mu , \hat{\tilde{x}}^\nu] = \tilde{\theta}^{\mu \nu} = \left( \begin{array}{cccc}
0 & 0 & 0 & \tilde{\theta^{03}} \\
0 & 0 & \tilde{\theta^{12}} & 0 \\
0 & -\tilde{\theta^{12}} & 0 & 0 \\
-\tilde{\theta^{03}} & 0 & 0 & 0
\end{array} \right).
\end{equation}
From here on we use these coordinates and drop the tilde. We can then define two sets of creation and annihilation operators:
\begin{equation}
\hat{b} = \frac{1}{\sqrt{2 \theta^{12}}}(\hat{x} + i\hat{y}) , \hat{b}^\dagger = \frac{1}{\sqrt{2 \theta^{12}}} (\hat{x} - i\hat{y}),
\end{equation}
\begin{equation}
\hat{a} = \frac{1}{\sqrt{2 \theta^{03}}}(\hat{t} + i\hat{z}) , \hat{a}^\dagger = \frac{1}{\sqrt{2 \theta^{03}}} (\hat{t} - i\hat{z}).
\end{equation}
The configuration space is then simply the two boson Fock space:
\begin{equation}
\mathcal{H}_c = \mbox{span} \{ \ket{n,m} = \frac{1}{\sqrt{n!m!}} (a^\dagger)^n (b^\dagger)^m \ket{0,0}  \}_{n,m=0}^{n,m = \infty}.
\end{equation}
Here the $c$ denotes this as the configuration Hilbert space and the adjoint on this space is denoted by $\dagger$.  As discussed above, we identify the algebra of fields with the Hilbert space of Hilbert-Schmidt i.e.:
\begin{eqnarray} \label{hq}
\mathcal{H}_q = \{ \hat{\psi} : \Tr_c(\hat{\psi}^\dagger \hat{\psi}) < \infty \}.
\end{eqnarray}
We denote the states in this space by $|\psi)$. This space has a natural inner product  $(\phi|\psi) = \Tr_c(\hat{\phi}^\dagger \hat{\psi}) $. The adjoint on this space is denoted by $\ddagger$ and, in analogy with ordinary quantum mechanics, we use the subscript $q$ to distinguish it from configuration space ${\cal H}_c$ and to indicate this as the space of quantum fields, which we shall refer to as the quantum Hilbert space.

We still need to define the notion of a derivative on these operator valued fields. This can be done in a purely algebraic fashion:
\begin{equation} 
\hat{\partial}_\mu \hat{\psi} = - i (\theta^{-1})_{\mu \nu} \left[\hat{x}^\nu , \hat{\psi} \right] .
\end{equation}
One can trivially check that this definition respects the Leibniz rule, which is one of the fundamental algebraic properties of a derivative.  One may be concerned that this definition only works when ${\rm det}\,\theta\ne 0$.  However, if ${\rm det}\,\theta=0$, it is always possible to make a change of variables such that there is at least one commuting variable.  The derivatives in the commuting variables are then defined as usual.  This procedure is discussed in detail in \cite{sinha}.  

If we consider the trace of a derivative of a Hilbert Schmidt operator, then:
\begin{eqnarray}
\Tr(\hat{\partial}_\mu \hat{\psi}) & = -i (\theta^{-1})_{\mu \nu} \left(\Tr(\hat{x}^\nu\hat{\psi}) - \Tr( \hat{\psi}\hat{x}^\nu)\right) \nonumber \\
& = 0,
\end{eqnarray}
where (\ref{hq}) was used.  One may be concerned that the condition that $\hat\psi$ is Hilbert-Schmidt may not be sufficient to ensure that the individual traces above exists and cancel.  One can, however, also verify this result by writing the trace as an integral over a coherent state basis.  It then becomes the integral of the derivative of the symbol of $\hat\psi$ in the coherent state basis.  The latter vanishes as the Hilbert-Schmidt condition ensures that the symbols are square integrable.  The commutative analogue of this result is to integrate a total derivative and if the function vanishes at infinity, this integral also vanishes. This also clarifies the role of the condition that the fields are Hilbert-Schmidt, i.e (\ref{hq}), namely it corresponds to the requirement that the functions vanish at infinity and that there are no boundary terms. With this result in place we can also derive the non-commutative analogue of partial integration:
\begin{equation}
\Tr(\hat{\partial}_\mu \hat{\psi} \hat{\phi}) = - \Tr(\hat{\psi} \hat{\partial}_\mu \hat{\phi}).
\end{equation}
It follows easily from this that $i \hat{\partial}_\mu$ is hermitian on ${\cal H}_q$.

\section{Lorentz Symmetry} 
\label{Lorentz Symmetry}

Now that we have identified the space of operator valued fields and defined derivatives on it, we can investigate the representations of the Lorentz group carried by this space. Consider an infinitesimal spacial Lorentz transformation parametrized by $\omega^\mu$. Adopting the same transformation properties for the non-commutative coordinates as for the commutative ones, we have:
\begin{equation}
\hat{x}^\mu \rightarrow  \hat{x}^\mu + \omega^\mu_\nu \hat{x}^\nu .
\end{equation}
Now consider a Lorentz scalar function transforming:
\begin{equation}
\hat\psi (\hat{x}^\mu) \rightarrow  \hat\psi (\hat{x}^\mu + \omega^\mu_\nu \hat{x}^\nu).
\end{equation}
To write this as a differential operator acting on $\hat{\psi}$, consider the Weyl representation \cite{doug}:
\begin{eqnarray}
\hat\psi(\hat{x}^\mu + \omega^\mu_\nu \hat{x}^\nu) &= \int d^4 k \, \phi(k) \mbox{exp} \left( i k_\mu (\hat{x}^\mu + \omega^\mu_\nu \hat{x}^\nu) \right) \nonumber \\
&= \int d^4 k \, \phi(k) (1 + \omega^\mu_\nu \hat{x}^\nu \hat{\partial}_\mu  - \frac{i}{2}  \omega^\lambda_\nu \theta^{\nu \mu} \hat{\partial}_\mu \hat{\partial}_\lambda ) \mbox{exp} \left( i k_\mu \hat{x}^\mu \right). \nonumber \\
\end{eqnarray}
The generators of the non-commutative Lorentz transformations are therefore given by \cite{Wess}:
\begin{equation}
\hat{J}_\omega = \hat{x}^\nu \omega^\mu_\nu \hat{\partial}_\mu  - \frac{i}{2}  \omega^\lambda_\nu \theta^{\nu \mu} \hat{\partial}_\mu \hat{\partial}_\lambda.
\end{equation}
Since
\begin{equation}
[\hat{J}_\omega,\hat{J}_{\omega^\prime}] = \hat{J}_{\omega \times \omega^\prime},
\end{equation}
where $\omega \times \omega^\prime = \omega^\rho_\nu {\omega^\prime}^\mu_\rho - {\omega^\prime}^\rho_\nu \omega^\mu_\rho,$ this satisfies the standard commutation relations of the Lorentz algebra and therefore constitute a valid representation, but in this case on a different Hilbert space. The coordinates transform in the expected way, like covariant $4$ vectors in commutative space: 
\begin{equation}
\hat{J}_\omega \hat{x}^\mu = \omega^\mu_\nu \hat{x}^\nu,
\end{equation}
and  $\hat{\partial}^\mu$ also transform like covariant 4 vectors since
\begin{equation}
[\hat{J}_\omega, \hat{\partial}^\mu] = \omega^\mu_\nu \hat{\partial}^\nu.
\end{equation}

\section{The covariant Dirac equation}
\label{Dirac}
\subsection{Free Dirac equation}

The free non-commutative Dirac equation takes the form, which we now know to be Lorentz invariant from the above:
\begin{equation}
(i \hat{\slashed{\partial}} - m) \hat{\psi} = 0.
\end{equation}
Inserting the operator valued versions of the commutative solutions in the non-commutative Dirac equation, for positive energy, where the $\chi$ are now eigenvectors of $\sigma^3$, with eigenvalues $s = \pm 1$ yields:
\begin{eqnarray} \label{free}
(i \hat{\slashed{\partial}} - m) e^{ -i k^\mu \hat{x}_\mu }
\frac{\slashed{k} + m}{\sqrt{2m(m + k^0)}}
\left( \begin{array}{cc}
\chi_s \\
0 
\end{array} \right) 
= e^{ -i k^\mu \hat{x}_\mu } \frac{(\slashed{k} - m)(\slashed{k} + m)}{\sqrt{2m(m + k^0)}}
\left( \begin{array}{cc}
\chi_s \\
0 
\end{array} \right). \nonumber \\ 
\end{eqnarray}
Requiring this to vanish yields the normal commutative free particle dispersion relation:
\begin{equation}
(k^0)^2 = E^2 = \vec{k}^2 + m^2.
\end{equation} 
Negative energy solutions can be constructed in a similar way or by using the C,P and T symmetries described further below. 

This outcome has already been obtained for many non-commutative theories, that is the free solutions retain the same form as commutatively and the spectrum is the same, the 2-D Schr\"odinger case can be found in \cite{scholtz1}. A different, but also valid solution has the exponential $e^{-i k^0 \hat{t}}  e^{-i k^1 \hat{x}} e^{-i k^2 \hat{y}} e^{ -i k^3 \hat{z}}$ , its various orderings are also solutions of (\ref{free}) only differing by a global phase.

\subsection{Lorentz covariant interactions}

To introduce interactions we must introduce operator valued fields, transforming appropriately under Lorentz transformations, into the Dirac equation.  However, if these fields act on the operator valued Dirac field in the sense of operator multiplication, the Lorentz symmetry is no longer manifest \cite{chaichian}.    The reason for this is that the Lorentz generators derived in Section \ref{Lorentz Symmetry} do not obey the Leibniz rule for operator multiplication:  
\begin{equation} 
\hat{J}_\omega ( \hat{\psi} \hat{\phi} ) = (\hat{J}_\omega \hat{\psi}) \hat{\phi} + \hat{\psi}(\hat{J}_\omega \hat{\phi}) - \frac{i}{2}( \theta^{\nu \rho} \omega^\mu_\nu - \theta^{\nu \mu} \omega^\rho_\nu) (\hat{\partial}_\mu \hat{\psi}) (\hat{\partial}_\rho \hat{\phi}).
\end{equation}
To make the Lorentz symmetry manifest, it is necessary to multiply operator valued fields differently to the normal operator product.   Consider products of the form:
\begin{equation}
\label{covprod}
\hat{\psi} \hat{*} \hat{\phi} \quad \mbox{where} \quad \hat{*} = \mbox{exp}(-\frac{i}{2} \theta^{\mu \nu} \hat{\overleftarrow{\partial_\mu}} \hat{\overrightarrow{\partial_\nu}}).
\end{equation}
This is similar to the Moyal $*$ star product used to combine ordinary functions to induce non-commutativity. However, (\ref{covprod}) differs by a sign and includes operator derivatives. Also this product is introduced for a different reason, not to induce non-commutativity, but to preserve Lorentz symmetry, it is as if one is undoing the regular Moyal star product. Combining fields in this way preserves Lorentz covariance since it can be readily shown that:
\begin{equation}
\hat{J}_\omega (\hat{\psi} \hat{*} \hat{\phi})  = (\hat{J}_\omega \hat{\psi}) \hat{*} \hat{\phi} + \hat{\psi} \hat{*} (\hat{J}_\omega \hat{\phi}). 
\end{equation}
This can also be thought of as modifying the multiplication map for operator products to untwist the deformed co-product.  Under this product operator multiplication actually becomes commutative as one can show that 
\begin{equation}
\hat{\psi} \hat{*} \hat{\phi} =\hat{\phi} \hat{*} \hat{\psi}.
\end{equation}

One might be concerned that this completely eradicates the effect of non-commutativity, particularly in light of the findings in \cite{oeckl} and \cite{bal}.  In \cite{oeckl} it was shown that quantum field theories on non-commutative $R^d$ are dual to quantum field theories on commutative $R^d$ with braided statistics, while in \cite{bal} it was argued that the $S$-matrix exhibits no dependence on the non-commutative parameter if the statistics is also twisted.  These results are, however,  concerned with the issue of twisting, which is of relevance to the multi-particle sector as was emphatically pointed out in \cite{bal}.    As we explicitly demonstrate below, it turns out that the modification implied by the use of the product (\ref{covprod}) leads to non-local effects that are already manifest on the one particle level where the issue of twisting is irrelevant.  These effects have clear physical implications, as discussed in section \ref{Landau}, where it is demonstrated that in the presence of a constant magnetic field non-commutative effects manifest themselves explicitly in the Aharonov-Bohm effect.  Assuming standard statistics, these single particle effects manifest themselves in the thermodynamics of Fermi systems as was extensively discussed in the non-relativistic case in \cite{scholtz4,scholtz5, scholtz6}.  In the case of the constant magnetic field discussed here, one expects that it will manifest itself in the filling fractions at incompressibility as is further discussed in section \ref{Landau}.

Finally, we also note that the invariant distance $s^2$ remains the same:
\begin{eqnarray}
s^2= \hat{x}^\rho \hat{*} \hat{x}_\rho = \hat{x}^\rho \hat{x}_\rho -\frac{i}{2} \theta^\rho_\rho = \hat{x}^\rho \hat{x}_\rho,
\end{eqnarray}
where the anti-symmetry of $\theta$ has been used.

\subsection{Electromagnetic Interactions}

Applying the above mentioned argumentation, we can now write down the covariant form of the non-commutative Dirac equation for particles coupled to a Lorentz covariant electro-magnetic field:
\begin{equation} \label{elec}
(i \hat{\slashed{\partial}} - q \hat{\slashed{A}} \hat{*} - m) \hat{\psi} \equiv {\slashed D} \hat{\psi}= 0.
\end{equation}
Here $q$ is the charge, and $\hat{A}$ is the real four potential i.e. $\hat{A} = \hat{A}^\dagger$, the same choice that is made commutatively. 

It is useful to verify explicitly that this equation is covariant.  The full Lorentz generators that generate coordinate and spinor transformations are given by $\hat{S}_\omega + \hat{J}_\omega$. Here $\hat{J}_\omega = \hat{x}^\nu \omega^\mu_\nu \hat{\partial}_\mu  - \frac{i}{2}  \omega^\lambda_\nu \theta^{\nu \mu} \hat{\partial}_\mu \hat{\partial}_\lambda$, which transforms the coordinates as previously calculated. ${S}_\omega =  \frac{i}{4} \sigma^{\mu \nu} \omega_{\mu \nu}$ gives the spinorial transformation properties of $\hat{\psi}$, which we take to be the same as commutatively since the spinor structure is not changed by non-commutativity. Here $ \sigma^{\mu \nu} = \frac{i}{2}[\gamma^\mu,\gamma^\nu]$. The Dirac equation is therefore covariant as it transforms in the same way as $\hat{\psi}$, since:
\begin{equation}
[(i \hat{\slashed{\partial}} - q \hat{\slashed{A}} \hat{*} - m),\hat{S}_\omega + \hat{J}_\omega] = 0,
\end{equation}
which again motivates the use of $\hat{*}$. This calculation can be found in \ref{appendix1}.

As is well known, quantum electrodynamics is a $U(1)$ gauge invariant theory.  We therefore expect a similar gauge symmetry for the non-commutative, interacting Dirac equation introduced in (\ref{elec}).  However, as usual multiplication by unitary operators will break Lorentz symmetry, the non-commutative gauge transformation must be implemented in the following form:
\begin{equation}
\hat{\psi} \rightarrow \hat{U} \hat{*} \hat{\psi}.
\end{equation}
Here $\hat{U}$ must have the following property to leave the inner product on the space of operator valued fields invariant and thus be a gauge transformation:  
\begin{equation}
\hat{U}^\dagger \hat{*} \hat{U} = \mathbb{I}.
\end{equation}
Thus $\hat{U}$ has a modified unitarity condition, which involves $\hat{*}$ instead of regular operator multiplication. Similarly $\hat{A}^\mu$ must undergo the following transformation to leave the non-commutative Dirac equation invariant:
\begin{equation}
\hat{\slashed{A}} \rightarrow \hat{U} \hat{*} \hat{\slashed{A}} \hat{*} \hat{U}^\dagger + \frac{1}{q} i (\slashed{\partial} \hat{U}) \hat{*} \hat{U}^\dagger.
\end{equation}
This retains the same form as commutatively, but with $\hat{*}$ instead of regular products.

\subsection{Discrete symmetries}

\subsubsection{Parity:}

We choose the non-commutative spacial parity transformation $\hat{P}^{(0)}$, to transform $\hat{x}^i \rightarrow - \hat{x}^i$ and $\hat{x}^0 \rightarrow \hat{x}^0$, to match commutatively. The full Parity transformation for spinors is then $\hat{P} = \gamma^0\hat{P}^{(0)}$. Then the non-commutative Dirac equation with interactions (\ref{elec}) transforms in the same as commutatively if potentials are mapped $A(x) \rightarrow A(\hat{x}) \equiv \hat{A}$.

\subsubsection{Charge conjugation:}

Commutatively the spacial charge conjugation transform was produced by complex conjugation $\hat{K}^{(0)}\psi = \psi^*$, the transformation for spinors is given by $\hat{C} = i \gamma^2 \hat{K}^{(0)}$. The non-commutative analogue of complex conjugation, is the adjoint $\dagger$ on the configuration space. However, we cannot simply take $\dagger$ since we do not want to take the adjoint of the spinors, but only the adjoint of each component. Let $\hat{K}^{(0)} = \hat{\Theta} \hat{\mathbb{I}}_4$, where $\hat{\Theta} \hat{\phi} = \hat{\phi}^\dagger$, note $\hat{\Theta} \hat{\partial}^\mu \Theta^{-1} = \hat{\partial}^\mu$. Then 
\begin{equation}
 \hat{\Theta} \hat{\mathbb{I}}_4 \left( \begin{array}{c} \hat{\phi_1} \\ \hat{\phi_2} \\ \hat{\chi_1} \\ \hat{\chi_2} \end{array} \right) = \left( \begin{array}{c} \hat{\phi^\dagger_1} \\ \hat{\phi^\dagger_2} \\ \hat{\chi^\dagger_1} \\ \hat{\chi^\dagger_2} \end{array} \right),
\end{equation}
using this new $\hat{K}^{(0)}$ the non-commutative Dirac equation with interactions (\ref{elec}) transforms in the same way, under $\hat{C}$, as commutatively if potentials are mapped as follows $A(x) \rightarrow A(\hat{x}) \equiv \hat{A}$, however, now with $\hat{A}^\dagger$ instead of $A^*$. The $\hat{*}$ is necessary since $\hat{K}^{(0)}$ switches the order of $\hat{A},\hat{\psi}$

\subsubsection{Time Reversal:}

We choose the non-commutative spacial time reflection transformation $\hat{T}^{(0)}$, to transform $\hat{x}^i \rightarrow  \hat{x}^i$ and $\hat{x}^0 \rightarrow - \hat{x}^0$, to match commutatively. Spacial time reversal is then given by, $\hat{K}^{(0)}\hat{T}^{(0)}$, using the above. The full transformation for spinors is $\hat{T} = i \gamma^1 \gamma^3 \hat{K}^{(0)} \hat{T}^{(0)} $. Again the non-commutative Dirac equation with interactions (\ref{elec}) transforms in the same way, under $\hat{T}$, as commutatively if potentials are mapped as follows $A(x) \rightarrow A(\hat{x}) \equiv \hat{A}$, using $\hat{A}^\dagger$ instead of $A^*$.

\subsubsection{CPT:}

The non-commutative $CPT$ transformation is given by $\widehat{CPT} = i \gamma^5 \hat{P}^{(0)} \hat{T}^{(0)} = - \gamma^2 \gamma^0 \gamma^1 \gamma^3 \hat{P}^{(0)} \hat{T}^{(0)}$. From the above results we can see $\widehat{CPT}$ symmetry is conserved in the non-commutative setting. We note that $\widehat{CPT}$ symmetry is also conserved for potentials which are not multiplied with $\hat{*}$, these are in general not Lorentz invariant, since $\widehat{CPT}$ doesn't involve a $\hat{\Theta}$. However, these types of potentials will break $C,T$ individually, but of course, the combined $CT$ is still conserved. In all of the discrete symmetries we have chosen potentials $A(x) \rightarrow A(\hat{x}) \equiv \hat{A}$ this is reasonable since when we take the commutative limit we would like to recover $A(x)$.

\section{Constant background magnetic field}
\label{Landau}

For the case of a constant background non-commutative magnetic field in the positive $ z$-direction we choose the gauge potential:
\begin{equation}
\hat{A}^\mu = \left( \begin{array}{c}
0\\ -\hat{x}^2 B\\ 0\\ 0 \end{array} \right).
\end{equation}
This is inspired by the form of the commutative potential since in the limit when $\theta$ goes to zero we should recover this. Putting this into (\ref{elec}) and calculating the $\hat{*}$ contribution gives:
\begin{equation} 
\label{DiracLand}
(i \hat{\slashed{\partial}} - \gamma^1 B q(\hat{x}^2 - \frac{i}{2} \theta^{2 1}\hat{\partial_1}) - m) \hat{\psi} = 0.
\end{equation}
Setting $\hat{\psi} = \left( \begin{array}{c} \hat{\phi} \\ \hat{\chi} \end{array} \right) $ gives two equations for the electron and positron components,
\begin{eqnarray} 
(- i\sigma_j \hat{\partial_j}  + \sigma_1( B q \hat{x}^2 - B q \frac{i}{2} \theta^{2 1} \hat{\partial_1})) \hat{\chi} = (i \hat{\partial_0} - m) \hat{\phi},  \\
(- i\sigma_j \hat{\partial_j}  + \sigma_1( B q \hat{x}^2 - B q \frac{i}{2} \theta^{2 1} \hat{\partial_1})) \hat{\phi} = (i \hat{\partial_0} + m) \hat{\chi}.
\end{eqnarray}
Solving for one of the spinors gives:
\begin{equation}
(- i\sigma_j \hat{\partial_j}  +\sigma_1(B q \hat{x}^2 - B q \frac{i}{2} \theta^{2 1} \hat{\partial_1} ))^2 \hat{\phi} =  (-\hat{\partial_0}^2 - m^2) \hat{\phi} .
\end{equation}
We define the magnetic length as usual:
\begin{equation}
l_B = \frac{1}{\sqrt{B |q|}} ,
\end{equation}
introduce a new variable,
\begin{equation}
\hat{\xi} = \frac{1}{l_B}\left(\hat{x}^2 + \mbox{sgn}(q) l_B^2 k^1 - \frac{1}{2} \theta^{2 1} {k_1} \right)
\end{equation}
and set:
\begin{equation}
a_\pm = l_B^2 \left( (k^0)^2 - m^2 - (k^3)^2 \pm \mbox{sgn}(q) \frac{1}{l_B^2}\right).
\end{equation}
Taking $ \hat{\phi} $ to be eigenvectors of $\sigma_3$, and plane wave solution in the $x-z$ plane gives:
\begin{eqnarray}
&\hat{\phi}_+ = \left( \begin{array}{c} F_+(\hat{\hat{\xi}}) \\ 0 
\end{array} \right) \mbox{exp} \left( -i k^0 \hat{t} + i k^1 \hat{x}^1 + i k^3 \hat{x}^3 \right), \nonumber \\ 
&\hat{\phi}_- = \left( \begin{array}{c}
0 \\ F_-(\hat{\hat{\xi}}) \end{array} \right) \mbox{exp} \left( -i k^0 \hat{t} + i k^1 \hat{x}^1 + i k^3 \hat{x}^3 \right) .
\end{eqnarray}
We could have also chosen the ordering with the plane wave to the right of the spinor, then moving the exponentials all the way to the left produces a shift $\hat{y} \rightarrow \hat{y} + \theta^{1 2}k_1 $. So the equation for this $\hat{\phi}$ is the same but the sign of the $\theta$ terms changes. These solutions are both valid but not independent. 
The following equation for $F$ is obtained, 
\begin{equation}
\left( \frac{d^2}{d\hat{\xi}^2} - \hat{\xi}^2 + a_\pm \right) F_\pm(\hat{\xi}) = 0.
\end{equation}
The commutative version of this can be found in \cite{zuber} and \cite{Bhattacharya}. These solutions match the form of the commutative case except that we have a modified $\hat{\xi}$ parameter and these are operator valued functions, in particular we can see that in our case the effect of non-commutativity was to shift $k^1$. Solutions of this differential equation are operator valued Hermite functions,
\begin{equation}
F_{n_\pm}(\hat{\xi}) =  N_{n_\pm} e^{-\hat{\xi}^2/2} H_{n_\pm} (\hat{\xi}).
\end{equation}
Here $H$ is a hermite polynomial and $a_\pm = 2 n_\pm + 1 , n_\pm \in \mathbb{N}$. This gives for the energies:
\begin{eqnarray} 
E^2_{n_\pm} = & \frac{1}{l_B^2} (2 n_\pm + 1 \mp \mbox{sgn}(q)) + m^2 + (k^3)^2.
\end{eqnarray}
We see that exactly the same spectrum as in the commutative case is obtained, which is expected since the energy doesn't depend on $k^1$.   There is a discrete 2-fold degeneracy, e.g. when $q = -e$, then $E_{n_+ - 1} = E_{n_-}$.   In this case we can therefore label the energies by $n_- = n$ and thus $E_n = \frac{2 n}{l_B^2}  + m^2 + (k^3)^2 $ and we note that only the ground state is non-degenerate $E_{n_- = 0}$. There is also a continuous degeneracy in $k^1$, reflecting the well-known degeneracy in Landau levels.  These results are all in agreement with the commutative case so that non-commutativity has no observable effect at the level of the density of states.  If we take for the normalization:
\begin{equation}
N_n = \left(\frac{\theta^{1 2}}{2^n n!}\sqrt{\frac{1}{\pi l_B^2}}\right)^{\frac{1}{2}},
\end{equation}
we obtain:
\begin{eqnarray}
\hat{\chi}_{+,n} = \frac{1}{E + m} \left( \begin{array}{c}
k^3 F_{n_+}(\hat{\xi}) \\ (\mbox{sgn}(q)/l_B) \sqrt{2 n_+ + (1 - \mbox{sgn} (q))} F_{(n_+ - \mbox{sgn} (q))}(\hat{\xi}) \end{array} \right) \nonumber  \\
\hat{\chi}_{-,n} = \frac{1}{E + m} \left( \begin{array}{c}
(\mbox{sgn}(q)/l_B) \sqrt{2 n_- + (1 + \mbox{sgn} (q))} F_{(n_- + \mbox{sgn} (q))}(\hat{\xi})  \\ - k^3 F_{n_-}(\hat{\xi}) \end{array} \right).
\end{eqnarray}

These are, however, not the most general solutions. Since there is a degeneracy in $k^1$, we can integrate with some function $f(k^1)$ i.e. $\int d k^1 f(k^1) \psi$, and still obtain an energy eigenvalue. We can choose $f(k^1) = e^{- \alpha^2 (k^1)^2}$, instead of the $\delta(k^1-(k^1)^\prime)$ for the plane-wave solution, this will make computation easier and introduces a localization in $\hat{x}^1$, determined by the length scale $\alpha^2$. The normalization of these states is given by $N_n =  \left( \frac{\theta^{1 2}}{2^n n!} \sqrt{\frac{2 \alpha^2}{\pi} \frac{1}{\pi l_B^2}} \right)^{\frac{1}{2}} $.

It can be readily shown by calculating $\Delta x \Delta y $, no matter what choice of $\alpha$ or $B$ we make, we will never obtain a localization below the length scale $\theta^{1 2}$, unlike commutative physics.  This is already a physical manifestation of non-commutativity as it implies an incompressible behaviour regardless of the value of the magnetic field.  Non-commutativity also manifests itself in the Aharonov-Bohm phase, which is physically observable through interference effects.

To derive the Aharonov-Bohm phase we need to calculate the phase difference that occurs when a state is transported around a closed loop. To do this we need the correct translation operators, which involve the usual translations, but also a gauge transformation since the Dirac operator (\ref{DiracLand}) is not manifestly translation invariant.  Commutatively we know these are $e^{i \vec{\Delta}\cdot\vec{\Pi}}$  for a spacial translation $\vec{\Delta}$,  where $\vec{\Pi}$ are the pseudo-momentum or magnetic translations, which commute with the commutative Hamiltonian \cite{Yoshioka}. Explicitly our pseudo-momentum are:
\begin{eqnarray}
& \hat{\Pi}_1 = i \hat{\partial}_1, \\
& \hat{\Pi}_2 = i \hat{\partial}_2 - \hat{x}^1 \frac{B q}{1 + \frac{B q \theta^{1 2}}{2}}, \\
& \hat{\Pi}_3 = i \hat{\partial}_3. 
\end{eqnarray}
These operators are different from the commutative case, but commute with the Dirac operator (\ref{DiracLand}) as shown in  \ref{appendix2} and are therefore the correct generators of translations. Transporting in a closed loop in the $x-y$ plane therefore yields
\begin{eqnarray}
e^{- i \hat{\Pi}_2 \Delta_2} e^{- i \hat{\Pi}_1 \Delta_1} e^{ i \hat{\Pi}_2 \Delta_2} e^{i \hat{\Pi}_1 \Delta_1} \hat{\psi} = \mbox{exp} \left( i \Delta_1 \Delta_2 \frac{B q}{1 + \frac{B q \theta^{1 2}}{2}} \right) \hat{\psi}.
\end{eqnarray}
The non-commutativity can be thought of as modifying the magnetic field:
\begin{equation}
B_\theta = \frac{B}{1 + \frac{B q \theta^{1 2}}{2}}.
\end{equation}
We can see an explicit physical effect caused by the non-commutativity. We know that for a confined system the degeneracy of a Landau level is given by the magnetic flux divided by the fundamental flux unit. If we consider the limit $B \rightarrow \infty$, then $B_\theta = \frac{2}{q \theta^{1 2}}$. So, even with an infinite magnetic field, the degeneracy is still finite since the non-commutativity restricts the maximum localization to a length scale $\theta$. This result also implies that the number of particles that can be accommodated in a Landau level is  modified by non-commutativity and correspondingly that incompressibility will occur at modified filling factors.  This is reminiscent of the fractional quantum Hall effect and offers the possibility of viewing interacting electrons moving in a constant magnetic field as an effective non-commutative theory of non-interacting electrons moving in the same magnetic field.  This point of view was already discussed in \cite{scholtz3}.

\section{Non-commutative action}
\label{action}

Our aim in this section is to construct an action on the operator level that yields the interacting, non-commutative Dirac equation (\ref{elec}) under variation.  Here we shall use the Hilbert space structure of the Hilbert-Schmidt operators explicitly by writing the action as an inner product in this space:
\begin{equation}
S[\hat{\psi}] = (\gamma^0 \hat{\psi}, \hat{\slashed{D}} \hat{\psi}) = \Tr_c(\hat{\bar{\psi}} \hat{\slashed{D}} \hat{\psi}).
\end{equation}
Next we note:
\begin{eqnarray}
\label{star}
& \Tr(\hat{\psi} \hat{*} \hat{\phi}) = \Tr(\hat{\psi} \mbox{exp}(-\frac{i}{2} \theta^{\mu \nu} \hat{\overleftarrow{\partial_\mu}} \hat{\overrightarrow{\partial_\nu}}) \hat{\phi}) \nonumber \\
& =\Tr(\hat{\psi} \mbox{exp}(\frac{i}{2} \theta^{\mu \nu} \hat{\overrightarrow{\partial_\mu}} \hat{\overrightarrow{\partial_\nu}}) \hat{\phi}) \nonumber \\
& =\Tr(\hat{\psi} \hat{\phi}),
\end{eqnarray}
which implies $\hat{*}$ doesn't change the inner product due to (\ref{hq}). So our action is Lorentz invariant and we can write the action as either $\Tr_c(\hat{\bar{\psi}} \hat{\slashed{D}} \hat{\psi})$ or $\Tr_c(\hat{\bar{\psi}} \hat{*} (\hat{\slashed{D}} \hat{\psi}))$.
Under a variation we obtain two conditions,
\begin{equation}
{\slashed{D}} \hat{\psi} = 0, \hat{\slashed{D}}^\ddagger \gamma^0 \hat{\psi} = 0.
\end{equation} 
The first is just the Dirac equation. To clarify the second condition we note that $(\hat{\slashed{A}\hat{*}})^\ddagger = \gamma^0 \hat{\slashed{A}\hat{*}} \gamma^0$, since:
\begin{eqnarray}
& (\hat{\phi},\hat{\slashed{A}}\hat{*}\hat{\psi}) = \Tr_c(\hat{\phi}^\dagger (\hat{\slashed{A}}\hat{*} \hat{\psi})) \nonumber \\
& = \Tr_c((\hat{\phi}^\dagger\hat{*}\hat{\slashed{A}})\hat{\psi}) \nonumber \\
& = \Tr_c((\hat{\slashed{A}}^\dagger \hat{*} \hat{\phi})^\dagger \hat{\psi}) \nonumber \\
& = (\gamma^0 \hat{\slashed{A}} \gamma^0 \hat{*} \hat{\phi}, \hat{\psi} ).
\end{eqnarray}
and we have used $(\hat{A} \hat{*} \hat{\psi})^\dagger = \hat{\psi}^\dagger\hat{*}\hat{A}^\dagger = \hat{A}^\dagger \hat{*} \hat{\psi}^\dagger $.

\subsection{The coherent state representation}

It is possible to define a four dimensional coherent state in configuration space \cite{Klauder}:
\begin{equation}
\ket{z,w} = e^{-\frac{|z|^2}{2}}e^{-\frac{|w|^2}{2}} e^{z \hat{a}^\dagger} e^{w \hat{b}^\dagger} \ket{0,0}.
\end{equation}
In the quantum Hilbert space we define a state, which is shown to have the interpretation as a minimum uncertainty state in \cite{scholtz1}:
\begin{equation}
| z, w ) = \ket{z,w}\bra{w,z},
\end{equation}
where:
\begin{equation} \label{zw}
z = \frac{1}{\sqrt{2 \theta^{03}}}(x^0 + i x^3) , w = \frac{1}{\sqrt{2 \theta^{12}}}(x^1 + i x^2),
\end{equation}
and $z,w$ are dimensionless numbers.  These states form an over complete basis for the configuration space and it is possible to write the identity as:
\begin{equation}
\mathbb{I}_c = \frac{1}{\pi^2} \int d z d z^* d w d w^* \ket{z , w} \bra{z , w}.
\end{equation}
Thus we can write the action as
\begin{equation} \label{actionsymbol}
S[\hat{\psi}] = \frac{1}{\pi^2} \int d z d z^* d w d w^*  \bra{z , w} \hat{\bar{\psi}} \hat{\slashed{D}} \hat{\psi} \ket{z , w},
\end{equation}
which simply means that we are integrating over the expectation values of the operator $ \hat{\bar{\psi}} \hat{\slashed{D}} \hat{\psi}$ in the coherent state basis, also known as the symbol of the operator.

On the quantum Hilbert space on the other hand, the identity can be written as 
\begin{equation}
\mathbb{I}_q = \frac{1}{\pi^2} \int d z d z^* d w d w^* |z , w) \mbox{exp} \left( \overleftarrow{\partial_z} \overrightarrow{\partial_{z^*}} + \overleftarrow{\partial_w} \overrightarrow{\partial_{w^*}}\right) (z , w|.
\end{equation}
The above product is known as the Vorus product, commonly denoted $*_V$, and written in terms of the dimensionful coordinates in (\ref{zw}) it reads:
\begin{equation}
*_V=\mbox{exp}\left(\frac{i}{2} \theta_V^{\mu \nu} {\overleftarrow{\partial_\mu}}{\overrightarrow{\partial_\nu}}\right),
\end{equation}
where:
\begin{equation}
\theta^{\mu \nu}_V = 
\left( \begin{array}{cccc}
- i \theta^{03} & 0 & 0 & \theta^{03} \\
0 & - i \theta^{12} & \theta^{12} & 0 \\
0 & - \theta^{12} & - i \theta^{12} & 0 \\
- \theta^{03} & 0 & 0 & - i \theta^{03}
\end{array} \right).
\end{equation}
This matrix differs from $\theta^{\mu \nu} = -i [\hat{x^\mu},\hat{x^\nu}]$ since it has on-diagonal terms and is not antisymmetric. To distinguish between the two, it is convenient to denote the latter by $\theta^{\mu \nu}_M$, since this corresponds to the matrix used in the Moyal product.

As these are minimum uncertainty  states, the real and imaginary parts of $z$ and $w$ are the closest to our conventional notion of space-time coordinates as obtained from a measurement.

We can explicitly write the action in this coherent state basis in terms of the dimensionful coordinates as:
\begin{eqnarray}
S[\psi]=&\frac{1}{\theta^{1 2} \theta^{0 3} \pi^2} \int d^4 x \bar{\psi}(x) \mbox{exp} \left( \frac{i}{2} (\theta_V - \theta_M)^{\mu \nu} {\overleftarrow{\partial_\mu}} 
{\overrightarrow{\partial_\nu}} \right) 
\times \nonumber \\ 
&\left((i \slashed{\partial} - q \slashed{A}(x) \mbox{exp} \left( \frac{i}{2} (\theta_V - \theta_M)^{\mu \nu} {\overleftarrow{\partial_\mu}} 
{\overrightarrow{\partial_\nu}} \right) 
 - m) \psi(x) \right).
\end{eqnarray}
Here we write the symbols as $\psi(x) \equiv (z,w | \psi ) = \bra{z,w} \hat{\psi} \ket{w,z}$. The same notation applies to  $A(x)$ and $\bar{\psi}(x)$. The $\partial$ are derivatives with respect to the dimensionful coordinates in (\ref{zw}). 

The star product that now appears is the one with matrix: 
\begin{equation}
(\theta_V - \theta_M)^{\mu \nu} = 
\left( \begin{array}{cccc}
- i \theta^{03} & 0 & 0 & 0 \\
0 & - i \theta^{12} & 0 & 0 \\
0 & 0 & - i \theta^{12} & 0 \\
0 & 0 & 0 & - i \theta^{03}
\end{array} \right).
\end{equation}
This differs from the non-commutative field theories that are conventionally written down, based on the Moyal product. Note that the presence of this product implies non-locality of the field theory. In the coherent state basis the Lorentz generators that act on symbols have the form:
\begin{eqnarray}
\bra{z,w} \hat{J}_w \hat{\psi} \ket{w,z} & = x^\nu \omega^\mu_\nu *_V \partial_\mu  \psi(x)  - \frac{i}{2}  \omega^\lambda_\nu \theta^{\nu \mu}_M \partial_\mu \partial_\lambda \psi(x) \nonumber \\
& = (x^\nu \omega^\mu_\nu \partial_\mu  + \frac{i}{2}  \omega^\lambda_\nu \theta^{\nu \mu}_{V - M} \partial_\mu \partial_\lambda) \psi(x).
\end{eqnarray}
By construction, and as can also be directly verified, these operators still close on the Lorentz algebra. The action in the coherent state basis is then invariant under these transformations, since the operator $ \hat{\bar{\psi}}\hat{*} \hat{\slashed{D}} \hat{\psi}$ on the operator level is by construction Lorentz invariant, while the $\hat{*}$ can be dropped under the trace as was shown in (\ref{star}).

From the generators above, the invariant distance in the coherent state basis $s^2$ must be given by the symbol:
\begin{eqnarray}
& \bra{z , w} \hat{x}^\rho \hat{*} \hat{x}_\rho \ket{z , w} =  x^\rho \mbox{exp} \left( \frac{i}{2} (\theta_V - \theta_M)^{\mu \nu} {\overleftarrow{\partial_\mu}} 
{\overrightarrow{\partial_\nu}} \right) x_\rho \nonumber \\ & = x^\rho x_\rho + \frac{i}{2} g_{\rho \mu}\theta^{\rho \mu}_{V-M} \nonumber \\
& = x^\rho x_\rho + \theta^{1 2},
\end{eqnarray}
which is the same as commutatively up to constant, which again reflects our inability to resolve different space-time points below the length scale theta.  This also implies that we are still working in Minkowski space.  

Another major difference from usual non-commutative field theories is the allowed functions $\psi(x)$.  The inner product on the operator level is mapped onto the following inner product on symbols:
\begin{equation}
(\phi|\psi) = \Tr_c(\hat{\phi}^\dagger \hat{\psi}) \rightarrow \frac{1}{\theta^{1 2} \theta^{0 3} \pi^2} \int d^4 x \phi(x)^* \mbox{exp} \left( \frac{i}{2} (\theta_V - \theta_M)^{\mu \nu} {\overleftarrow{\partial_\mu}} {\overrightarrow{\partial_\nu}} \right) \psi(x).
\end{equation}
Thus, if the Hilbert space is the Hilbert-Schmidt operators then the space of symbols is not simply $L^2$ but the space where the above inner product is finite. In Fourier space this inner product becomes:
\begin{equation}
(\phi|\psi) \propto \int d^4 k \phi^*(k) \psi(k) \mbox{exp} \left( \frac{\theta^{1 2}}{2}((k^1)^2 + (k^2)^2) + \frac{\theta^{0 3}}{2}((k^0)^2 + (k^3)^2) \right).
\end{equation}

The condition that the operator valued fields are Hilbert-Schmidt, which is required to ensure that the above action is finite, is actually more restrictive than the square integrability of functions, it enforces a smoothness on length scales smaller than theta.
As we can see in the above equation, for the inner product to be finite, the high momentum modes must be suppressed because of the positive theta exponential. The suppression of high momentum modes implies, through the Fourier transform, a smoothness on small lengths scales. 

\section{Conclusion}
\label{concl}
Here we proposed a covariant interacting, non-commutative Dirac equation on the operator level.  The key element of this construction is a modified operator product that ensures that the Leibnitz rule applies, despite the modified form of the Lorentz generators when implemented on operator space.  The free non-commutative Dirac equation was solved and exhibits no deviation from commutative behaviour.  The interacting Dirac equation in the presence of a constant magnetic field was also solved.  In this case no modification from commutative behaviour was found on the level of the density of states, but the system does exhibit a different physical behaviour in terms of the possible localisation of the wave-function, and thus incompressibility, as well as the Aharonov-Bohm phase.  A Lorentz invariant action was derived that yields the interacting, non-commutative Dirac equation under variation.  When represented as a field theory, this action differs from the commonly proposed non-commutative actions based on the Moyal product.  In particular a different, commutative, star product appears, the measure of the path integral seems to be different and the action is non-local.

\appendix

\section{Covariance of the Dirac equation} \label{appendix1}

The full generators for Lorentz transformations for spinors are given by:
\begin{equation}
\hat{J}_{\omega_{\frac{1}{2}}} =  \hat{S}_\omega + \hat{J}_\omega.
\end{equation}
Where $\hat{J}_\omega = \hat{x}^\nu \omega^\mu_\nu \hat{\partial}_\mu  - \frac{i}{2}  \omega^\lambda_\nu \theta^{\nu \mu} \hat{\partial}_\mu \hat{\partial}_\lambda$ this transforms the coordinates as previously calculated. ${S}_\omega =  \frac{i}{4} \sigma^{\mu \nu} \omega_{\mu \nu}$ gives the spinorial transformation properties of $\hat{\psi}$, which we take to be the same as commutatively since the spinor structure is not changed by non-commutativity. Here $ \sigma^{\mu \nu} = \frac{i}{2}[\gamma^\mu,\gamma^\nu].$ A commutative discussion can be found in \cite{zuber}. To calculate the action of $\hat{S}_\omega$, the following is needed:
\begin{eqnarray}
& [\omega_{\mu \nu} \sigma^{\mu \nu},\gamma^\lambda] = \omega_{\mu \nu} 2 i  (g^\lambda_\mu \gamma_\nu - g^\lambda_\nu \gamma_\mu) \nonumber \\
& = 4 i  (\omega^{\lambda \nu} \gamma_\nu).
\end{eqnarray}
Transforming the Dirac equation, first by $\hat{S}_\omega$:
\begin{eqnarray}
(-i \omega^\mu_\nu \gamma^\nu \hat{\partial_\mu} + q \omega^\mu_\nu \gamma^\nu \hat{A_\mu} \hat{*}) \hat{\psi}  + (i \hat{\slashed{\partial}} - q \hat{\slashed{A}} \hat{*} - m) \hat{S}_\omega \hat{\psi},
\end{eqnarray}
and then by $\hat{J}_\omega$:
\begin{eqnarray}
& (i  \gamma^\mu \omega^\nu_\mu \hat{\partial_\nu} - q  \gamma^\mu (\hat{J}_\omega \hat{A_\mu}) \hat{*}) \hat{\psi}  + (i \hat{\slashed{\partial}} - q \hat{\slashed{A}} \hat{*} - m) \hat{J}_\omega \hat{\psi} \nonumber \\
& = (i  \gamma^\mu \omega^\nu_\mu \hat{\partial_\nu} - q  \gamma^\mu \omega^\nu_\mu \hat{A_\nu} \hat{*}) \hat{\psi}  + (i \hat{\slashed{\partial}} - q \hat{\slashed{A}} \hat{*} - m) \hat{J}_\omega \hat{\psi},
\end{eqnarray}
since $\hat{A}$ transforms, by definition, like a 4-vector under coordinate transformations. Putting this together:
\begin{equation}
(\hat{S}_\omega + \hat{J}_\omega)(i \hat{\slashed{\partial}} - q \hat{\slashed{A}} \hat{*} - m) \hat{\psi} =  (i \hat{\slashed{\partial}} - q \hat{\slashed{A}} \hat{*} - m) (\hat{S}_\omega + \hat{J}_\omega) \hat{\psi}.
\end{equation}
The Dirac equation is therefore covariant as it transforms in the same way as $\hat{\psi}$. Note however the Dirac operator $(i \hat{\slashed{\partial}} - q \hat{\slashed{A}} \hat{*} - m)$ is Lorentz invariant. This implies that a Lorentz transformation transforms the solutions of the Dirac equation into each other. This provides an additional motivation for the $\hat{*}$ coupling, since otherwise the action of $\hat{J}_\omega$ would have produced an additional term as seen previously.

\section{Non-commutative magnetic translation operators} \label{appendix2}

Our translation operators are given by
\begin{eqnarray}
& \hat{\Pi}_1 = i \hat{\partial}_1, \\
& \hat{\Pi}_2 = i \hat{\partial}_2 - \hat{x}^1 \frac{B q}{1 + \frac{B q \theta^{1 2}}{2}}, \\
& \hat{\Pi}_3 = i \hat{\partial}_3.
\end{eqnarray}
We show these are indeed correct by calculating their commutators with the Dirac operator (\ref{DiracLand}). Clearly $\hat{\Pi}_1 \hat{\Pi}_3$ commutate. We show $\hat{\Pi}_2$ also commutates by calculating the commutator of $i \hat{\partial}_2 - \hat{x}^1 B q \hat{*}$, with the relevant terms in (\ref{DiracLand}):

\begin{eqnarray}
& \left[i (1 +  \frac{B q \theta^{1 2}}{2}) \hat{\partial}_2 - B q \hat{x}^1, i \hat{\partial}_1 - i \frac{B q \theta^{1 2}}{2} \hat{\partial}_1 - B q \hat{x}^2 \right] = \nonumber \\
& - i (1 +  \frac{B q \theta^{1 2}}{2}) B q [\hat{\partial}_2, \hat{x}^2] - i B q [\hat{x}^1, \hat{\partial}_1] \nonumber \\ & + i \frac{(B q)^2 \theta^{1 2}}{2} [\hat{x}^1, \hat{\partial}_1] + (B q)^2 [\hat{x}^1,\hat{x}^2] = \nonumber \\ & 0.
\end{eqnarray}
From this it follows all our translation operators commutate with the Dirac operator (\ref{DiracLand}).

\section*{Acknowledgements}

PHW gratefully acknowledges the financial support of the South African National Research Foundation.

\end{document}